\documentclass[aps,prb,twocolumn,showpacs,preprintnumbers,amsmath,amssymb]{revtex4}
\usepackage[T2A]{fontenc}

\usepackage{epsf,graphicx}
\begin{document}
\title{The effect of electronic entropy on temperature peculiarities of the frequency characteristics of two interacting anharmonic vibrational modes in $\beta-$Zr.}   
\author{V.Yu.~Trubitsin}
\email{tvynew@otf.pti.udm.ru}
\affiliation{Physico-Technical Institute, Ural Branch of RAS, 132 Kirov Str., 426001 Izhevsk, Russia}
\date{\today}
                                       
\begin{abstract}
A 2D temperature-dependent effective potential is calculated for the interacting longitudinal and transverse  $L-$phonons of $\beta$ zirconium in the frozen-phonon model. The effective potentials obtained for different temperatures are used for the numerical solution of a set of stochastic differential equations with a thermostat of the white-noise type. Analysis of the spectral density of transverse vibrations allows one to determine the temperature at which  $\beta$-Zr   becomes unstable with respect to the longitudinal $L-$vibrations. The obtained temperature value practically coincides with the experimental temperature of the  $\beta \to \alpha$  structural transition in zirconium. The role of electronic entropy in the  $\beta-$Zr stability is discussed.
 \end{abstract}
\pacs{ 63.20Ry, 05.10Gg, 63.20 Kr, 71.15Nc}
\maketitle

In recent years, structural transformations of zirconium induced by high temperatures and pressures have been intensively investigated both experimentally and theoretically.  
Considerable attention has been given to studying, at atmospheric pressure, the structural stability of the high- temperature $\beta-$phase and the peculiarities of its transition into an $\alpha-$phase existing at temperatures below $T=1136K$ \cite{Tonkov}. 
The commonly accepted view today is that the structural $\beta \to \alpha$ phase transition at atmospheric pressure is closely related to the softening of the transverse phonon with wave vector $k=\frac{1}{2}[110]$ ($N-$phonon) experimentally observed in $\beta-$Zr \cite{Heiming,Heiming1}. Theoretical arguments in favour of this viewpoint are mainly based on calculations of the effective potential for the N-phonon performed in the frozen-phonon model \cite{Chen-85}. From these calculations it follows that the effective potential of the N-phonon of Zr has a two-well shape, and, as a consequence, the phonon frequency squared in the harmonic approximation is negative. The imaginary phonon frequency points to the instability of $\beta-$Zr in the ground state. 

Using the perturbation theory formalism for anharmonic effects in crystals, it was shown\cite{Chen-85} that at high temperatures the $\beta$-phase of Zr becomes stable to the atomic displacements corresponding to the N-phonon due to its interaction with other transverse vibrational modes lying along the (110) direction. Using a modified pseudoharmonic approximation \cite{Salamatov-96}, we in turn succeeded in showing that in this case stability may be attained by merely allowing for the inherent anharmonicity of the N-mode \cite{Ostanin-98}.

However, the stability of $\beta-$Zr as a whole is not determined only by the lattice stability to the atomic displacements corresponding to the N-phonon.  The calculations performed in the "frozen"-phonon model \cite{Ho-84} show that $\beta-$Zr is also unstable with respect to the atomic displacements corresponding to the longitudinal vibrations with wave vector $k=\frac{2}{3}(1,1,1)$ ($L_l$-phonon). 

At pressures larger than the triple point ($T=973$~K and  $P=5.5$~GPa) such displacements cause the temperature phase transition from the $\beta$- to the $\omega$- phase (the $AlB_2$-type structure). At room temperature the $\omega$-phase of Zr is experimentally observed in the pressure range from 2.2 to 30-35 GPa \cite{Akahama1,Akahama2}. 
At zero pressure the instability of  $\beta-$Zr to this mode manifests itself in a sharp decrease of the longitudinal vibration frequency  obtained from the spectra of inelastic neutron scattering near $k=\frac{2}{3}(1,1,1)$  \cite{Heiming}.  To our knowledge none of the currently available theoretical studies discusses the mechanism of stabilization of the high-temperature $\beta-$phase with respect to the $L_l$ displacements. Moreover, it is still unclear whether stabilization is due to the inherent  anharmonicity of the $L_l$ mode or to the phonon-phonon interaction.

In Ref.~\onlinecite{Trubitsin-2004} we have suggested a model for two interacting modes with wave vector $k=\frac{2}{3}(1,1,1)$ (a strongly anharmonic longitudinal $L_l$ mode, and an almost harmonic transverse $L_t$ mode) embedded in a thermostat modelling their connection with the rest of the crystal. 

The temperature dependence of the vibration frequency of the $L_l$ and $L_t$ modes was found by solving a system of stochastic differential Langevin equations with a thermostat of the white-noise type, which describe the motion of a particle in a two-dimensional  effective potential $W(x,y)$. The potential was calculated within the electron density functional theory for a series of simultaneous atomic displacements $x,y$ corresponding to the $L_l$ and $L_t$ modes.
The interaction of these modes was shown to result in an induced anharmonicity for quick transverse $L_t$ vibrations which in the absence of interaction remain almost harmonic at all temperatures considered. The presence of induced anharmonicity leads to the frequency broadening of the spectral density $S(\omega)$ and to the appearance  of a complicated temperature dependence of $S(\omega)$ of the $L_t$ mode.

The calculation of the probability density of the mean-square atomic displacements \cite{Trubitsin-2004} shows that the probability of finding the atoms at the sites corresponding to the $bcc$ lattice ($\beta-$phase) increases with temperature. However, even at temperatures much higher than the point of the $\beta \to \alpha$ structural phase transition this probability is almost  three times smaller than the probability of finding the atoms at the $\omega$ sites. In other words, even at high temperatures the $bcc$ lattice remains unstable with respect  to the $L_l$ displacements. 

We have assumed \cite{Trubitsin-2004} the two-mode effective potential $W(x,y)$ to be temperature independent. This is a rough assumption because, firstly, it does not allow for the crystal lattice expansion with increasing temperature. Secondly, at finite temperatures the potential $W(x,y)$ should be calculated not from the total energy of a crystal in the ground state, $E_{el}$, but from the free energy $F(T)$, 
\begin{equation}
F(T) = E_{el}-TS_{el}, \label{Free_energy}
\end{equation}
where $S_{el}$ is the electronic entropy and $T$ is the temperature. In Ref.~\onlinecite{EWW} the electronic entropy was shown to be of considerable importance in stabilizing the high-temperature $bcc$ phase of Zr.

 The effect of the crystal volume change on the two-mode potential $W(x,y)$ will be detailed elsewhere. In this work we shall restrict ourselves to the calculation of the electronic entropy effect on $W(x,y)$. 
We shall also find the temperature dependence of the spectral density of the longitudinal, $L_l$, and transverse, $L_t$, modes with ${\bf k}=2/3(111)$ and determine, at zero pressure, the  stability region of the  $\beta$ phase of Zr with respect to the atomic displacements corresponding to these modes. To this end we shall successively consider the dependence of the electron entropy  $S_{el}(T,x,y)$ and free energy $F(T,x,y)$ on the temperature and atomic displacements $x,y$.  
The effective potentials $W_T(x,y)=F(T,x,y)-F(T,0,0)$ calculated at each fixed temperature will be used in solving a set of stochastic differential Langevin equations of motion with a white-noise thermostat. An analogous stochastic approach was earlier used \cite{Gornostyr-96} to model the lattice vibrations of a strongly anharmonic crystal.
In Ref.~\onlinecite{Gornostyr-96}, as distinct from this work, the dynamics of motion of a single longitudial $L_l$ mode in a one-dimensional temperature independent model potential $W(x)$ have been studied.                           

\section{The two-mode effective potential}
As in a previous paper \cite{Trubitsin-2004}, the unit cell of $\beta-$Zr was chosen as a hexagonal lattice with three base atoms. A detailed description of the unit cell geometry, its relation with the $bcc$ lattice parameters, and the atomic displacements corresponding to the vibrations with the chosen wave vector $k=2/3[111]$  may be found, for example, in Ref.~\onlinecite{Chen-85}. The total energy was calculated by the self-consistent full-potential LMTO  (FP LMTO)  method \cite{SAVRAS,SAVR} with  the GGA approximation for the exchange-correlation potential term \cite{GGA96}. 
The same set of FP LMTO parameters was used for all displacements. The one-centre expansions inside the MT spheres were confined to $l_{max}=6$ . The MT sphere radii were chosen equal to $R_{mt}=2.20$ a.u.  Integration over $k$ was performed on a (10,10,10) mesh equivalent to 166 $k$-points in the irreducible part of the Brillouin zone. The total energy was calculated  for 108 pairs of coordinates $x,y$ (18 values for the variable $x$ corresponding to the longitudinal $L_l$  mode, and 6 values for $y$ corresponding  to the  transverse $L_t$ mode). The energies obtained were first approximated by a tenth-degree polynomial for the longitudinal vibrations and then by a fourth-degree polynomial for the transverse ones. 

\begin{figure}[tbh]
\begin{center}
\resizebox{0.95\columnwidth}{!}{\includegraphics*{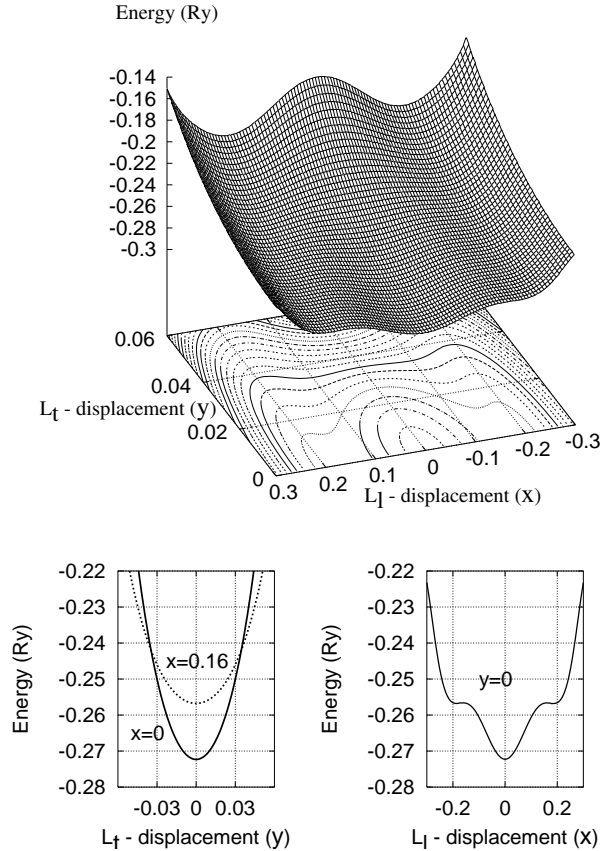}}
\caption{Two-mode effective potential for longitudinal and transverse atomic displacements corresponding to the vibrations with wave vector $k=\frac{2}{3}(1,1,1)$ in $\beta-$Zr.  The displacements are given in units of the lattice parameter $a$.}
\label{fig:Pot+Entropy}
\end{center}
\end{figure}

The two-mode effective potential calculated in the frozen-phonon model is shown in Fig.\ref{fig:Pot+Entropy} (the potential part corresponding to the positive period of transverse vibrations is presented in the upper panel, while the lower panels show the potential cross-sections for the transverse ($y=0$) and longitudinal ($x=0$)  vibration components.)  As seen from the figure, the total energy is minimum at zero displacements (the atoms are localized at the $\omega$  sites). Two local energy minima at $x \approx |0.166|, y=0$ correspond to the $bcc$ atomic arrangement. It should be noted that  to obtain the  $\omega$ structure observed in zirconium from the  $bcc$  lattice it does not suffice to merely  displace the atoms along the $x$-direction, one should also increase the  $c/a$ ratio. Nevertheless, in this paper we shall use the term "$\omega$ phase" for a lattice with zero displacements. At  $y=0$, the effective potential for longitudinal vibrations has a three-well shape, which agrees with the calculations in Ref.~\onlinecite{Chen-85}. Since the local minima are shallow, the $bcc$ lattice of Zr is unstable with respect to small longitudinal vibrations. At large  transverse displacements, the potential for the longitudinal component transforms into a two-well one.

The left lower panel of Fig.\ref{fig:Pot+Entropy} shows the cross-sections corresponding to the "pure" transverse atomic vibrations near the $\omega$  (solid line) and  $\beta$ (dotted line) centres. The effective potential is seen to have in either case  a parabolic shape, hence, the vibrations frequency for the transverse branch, in the absense of interaction with the longitudinal vibrations, must be described quite well by the harmonic approximation in both the $\omega$ and $\beta$ phases.

\section{Calculation of the electronic entropy}

The electronic exñitation entropy term in (1) will be introduced as earlier  
\cite{EWW}. This means that our analysis is limited to the classical 
lattice-dymanics regime when the electron-phonon interaction effects become 
negligible and the electronic excitation entropy term $S_{el.}$ is bare 
electronic entropy 
\begin{equation} 
S_{e} = - k_B \int { n [ f ln f - (1-f) ln (1-f)] dE} .
\end{equation} 
Here $n(E)$ is the electron density of states (DOS), and $f(E)$ is the Fermi distribution
\begin{equation} 
f(E) = {[\exp{\beta (E - \mu)} +1 ] }^{-1},
\end{equation} 
$\beta = 1/(k_B T)$. The chemical potential $\mu (T)$  is determined from the condition for 
the number of electrons $z$
\begin{equation} 
z =  \int { n (E)  f(E) dE}.
\label{Eq:him_potencial}
\end{equation} 
In paper \onlinecite{EWW} $S_{el} $ was calculated for the $hcp$ and $bcc$ structures of 
Zr and Ti. The electronic entropy was shown to be large and strongly 
nonlinear in temperature, as a result of both the volume and energy 
dependences of the DOS.

\begin{figure}[!tbh]
\begin{center}
\resizebox{0.95\columnwidth}{!}{\includegraphics*{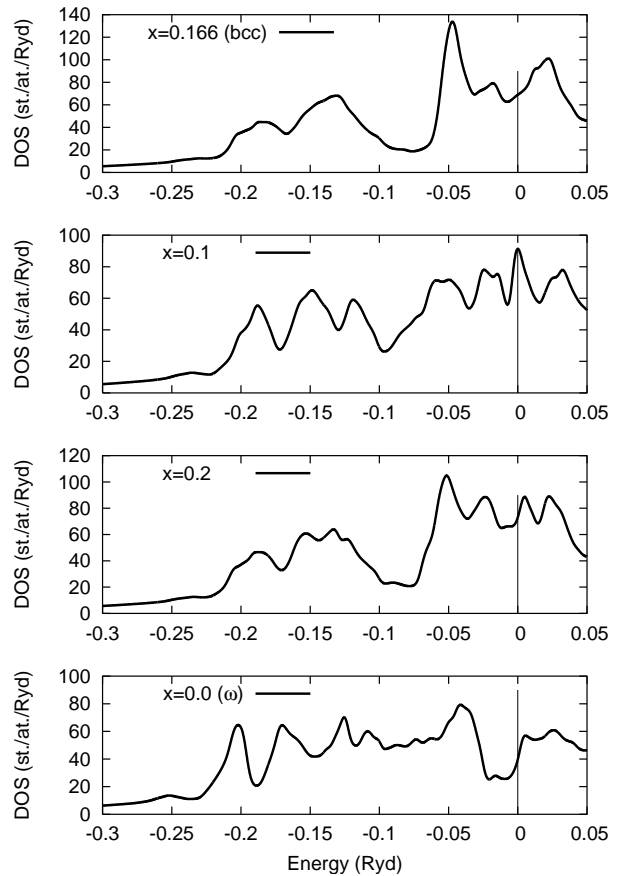}}
\caption{The DOS of Zr at different atomic displacements $x$ corresponding to the  $L_l$ vibrations. Zero displacements correspond to the  $\omega$-phase. The energy is reckoned  from the Fermi energy.}
\label{fig:L_DOS}
\end{center}
\end{figure}

In Fig.~\ref{fig:L_DOS},  $n(E)$  is shown for different atomic displacements corresponding to the  $L_l$
vibrations in the absence of transverse displacements ($y=0$). 

The bottom picture corresponds to the $\omega$ lattice ($x=0.0$), and the top one to the $bcc$ lattice ($x=0.166$).
It can be seen that a considerable change in the electron spectrum occurs at longitudinal atomic displacements. So,  the DOS at the Fermi level, $n(E_F)$, gets almost doubled on transition from the  $\omega$  to the $bcc$  structure. The DOS at the Fermi level reaches its maximum at $x=0.1$, when the Fermi level passes through the peak of DOS.

 \begin{figure}[tbh]
\begin{center}
\resizebox{0.95\columnwidth}{!}{\includegraphics*{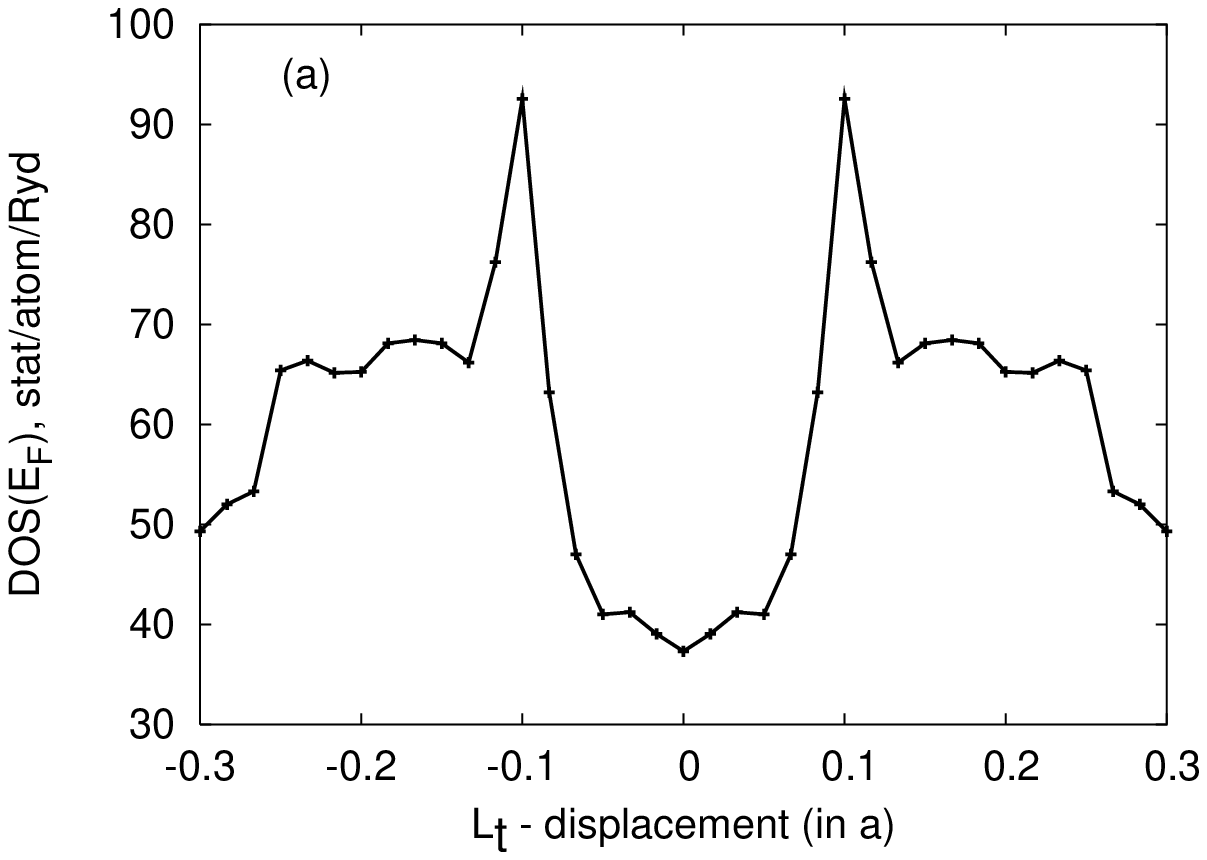}}
\end{center}
\begin{center}
\resizebox{0.95\columnwidth}{!}{\includegraphics*{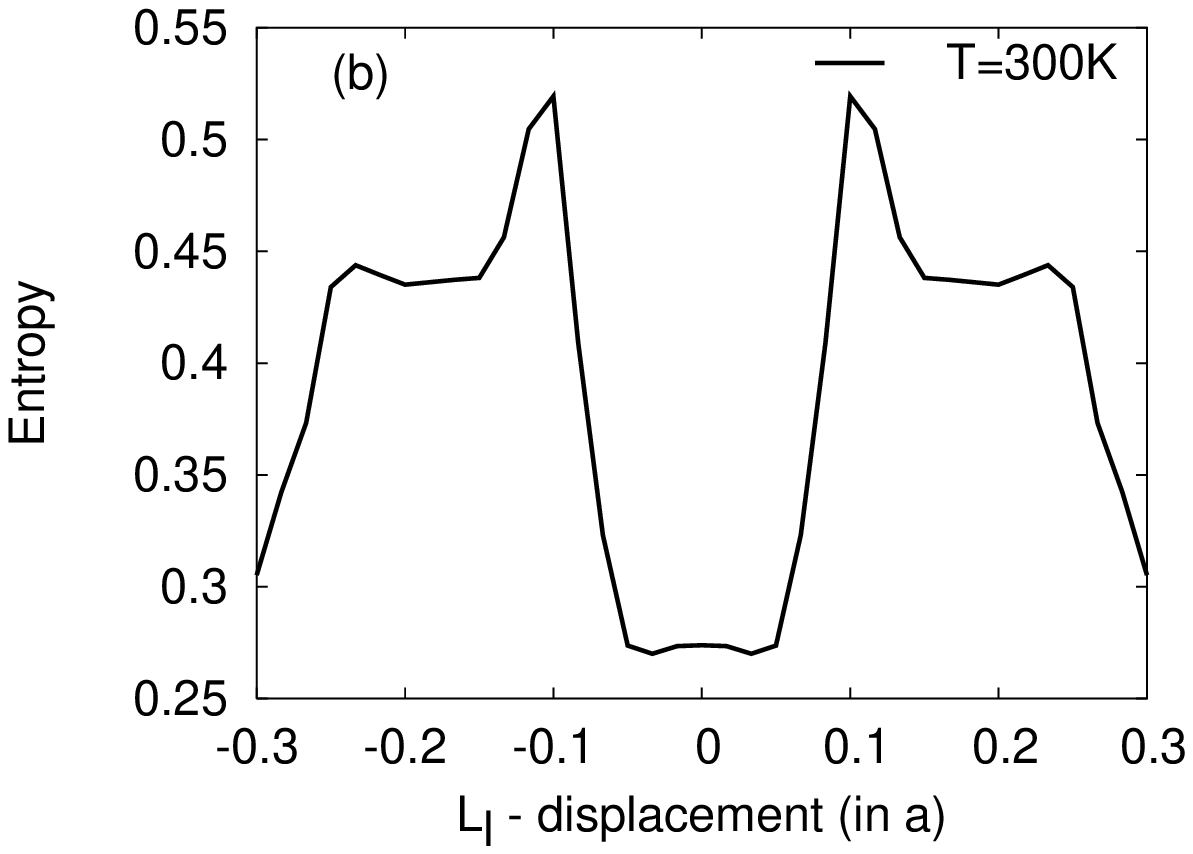}}
\caption{The DOS at the Fermi level (a) and entropy at  $T=300$K (b) for the atomic displacements corresponding to the  $L_l$ vibrations.}
\label{fig:Dos_on_ef_Entrop}
\end{center}
\end{figure}

The electronic DOS at the Fermi level, $n(E_F)$, and the entropy at $T=300$K calculated for the displacements corresponding to the   $L_l$ vibrations in  $\beta-$Zr are presented in Fig.\ref{fig:Dos_on_ef_Entrop}. It is seen that at room temperature the electronic entropy changes at $L_l$  displacements in the same way as the DOS at the Fermi level. This result is in good agreement with the expression for the entropy of a strongly degenerate electron gas at low temperatures, known from the electron theory of metals \cite{Lifshic}

\begin{equation}
S_e= \frac{\pi^2}{3} n(E_F) T.
\label{Eq:entrop_free_el_gaz}
\end{equation}
 
\begin{figure}[!tbh]
\begin{center}
\resizebox{0.90\columnwidth}{!}{\includegraphics*{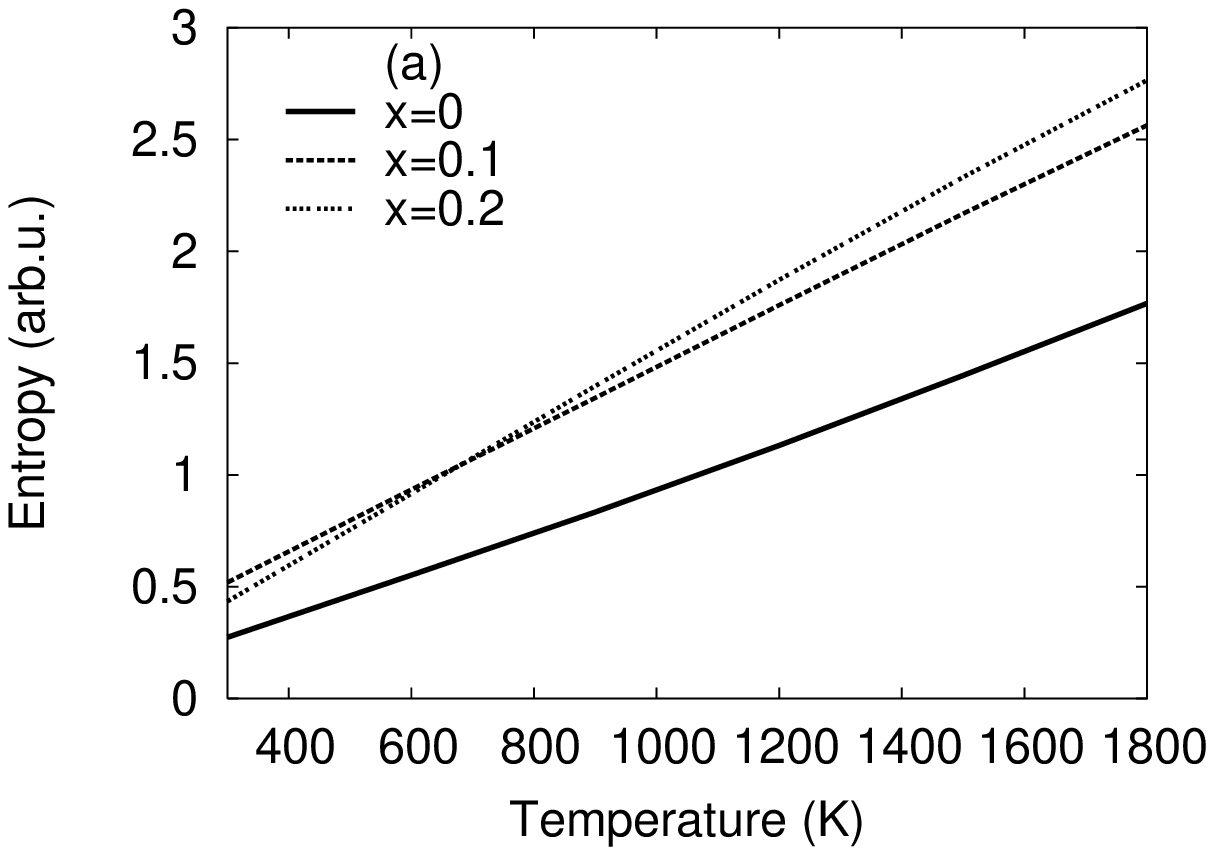}}
\resizebox{0.90\columnwidth}{!}{\includegraphics*{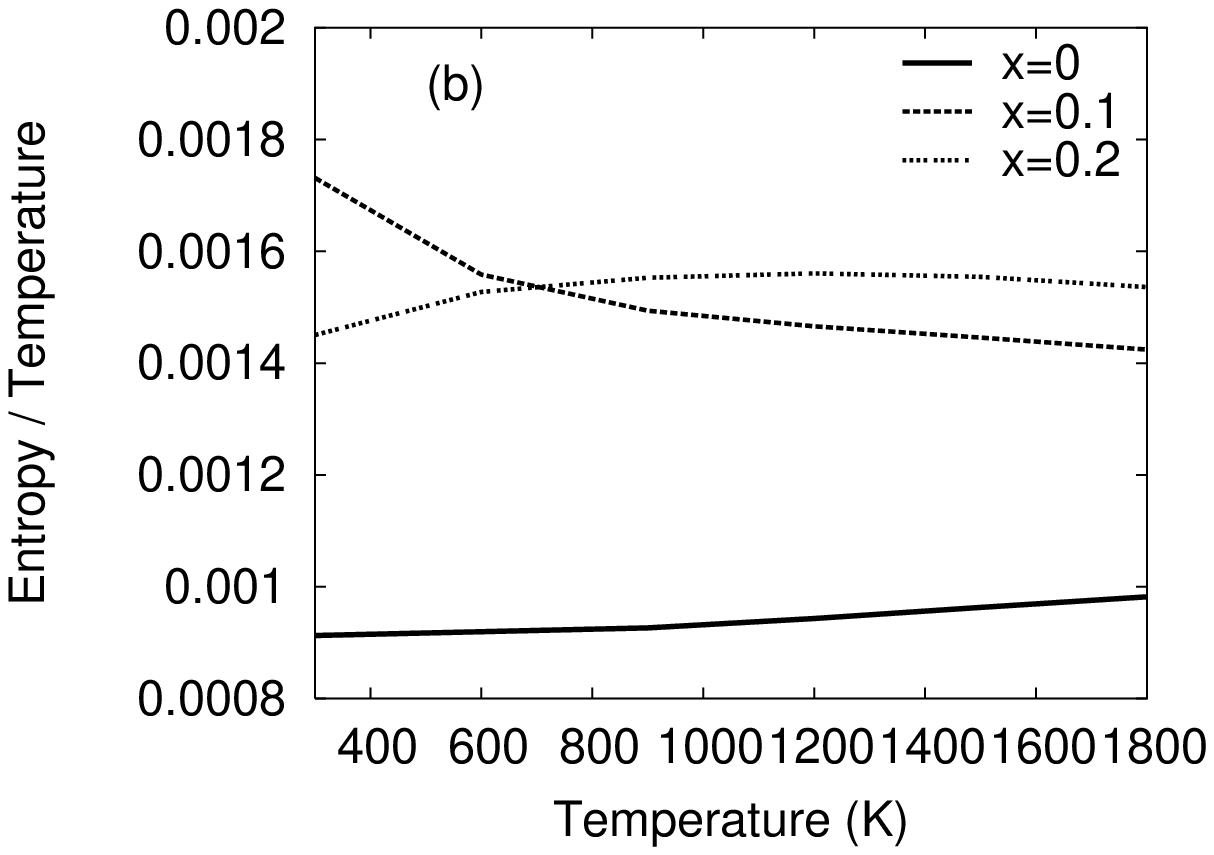}}
\caption{The temperature dependence of entropy (a), and the coefficient of proportionality between entropy and temperature (b) for different atomic displacements ($x$=0, 0.1 and 0.2).}
 \label{fig:Entrop_by_T}
\end{center}
\end{figure}

 From this expression it follows that, at a constant temperature, the entropy is proportional to the DOS at the Fermi level. Nonetheless, it can be seen from the figure that near $x=0$ and $x=0.16$ this proportionality is violated. To clarify the situation, let us take a look at Fig.\ref{fig:Entrop_by_T}.
 
In Fig. \ref{fig:Entrop_by_T}(a) the temperature dependence of the entropy is calculated for three atomic displacements corresponding to: (i) the $\omega$  lattice($x=0.0$), (ii) the maximum DOS value   $x=0.10$, and (iii) $x=0.20$. It is seen that, on the whole, in the temperature range considered the entropy is fairly well approximated by a linear function of temperature, which is in good agreement with the expression for the entropy of a strongly degenerate free electron gas (\ref{Eq:entrop_free_el_gaz}). Fig.\ref{fig:Entrop_by_T}(b) shows, however, that the coefficient of proportionality between entropy and temperature has an intricate temperature dependence. So,  it increases with temperature at $x=0.0$, and diminishes with increasing temperature at  $x=0.10$, while at $x=0.20$ the coefficient grows with a rise in temperature up to $T=1200K$ and decreases with further increase of $T$.  In relation (\ref{Eq:entrop_free_el_gaz}) the coefficient of proportionality between entropy and temperature is merely the DOS at the Fermi level. The temperature dependence of  $S/T$  in Fig.\ref{fig:Entrop_by_T}(b) can be obtained by replacated the  Fermi energy in relation (\ref{Eq:entrop_free_el_gaz}) for the chemical potential determined from the normalizing condition (\ref{Eq:him_potencial}). Indeed,  as seen in Fig.\ref{fig:L_DOS}, the DOS grows with increase in chemical potential for $x=0.0$, and decreases for $x=0.10$. The intricate shape of the $S/T$ curve at $x=0.20$ is due to the fact that with increasing temperature  $n(\mu(T))$  passes through the maximum. Correspondingly, entropy grows at low temperatures and diminishes at high ones. Thus, expression (\ref{Eq:entrop_free_el_gaz}) may  be used to calculate the entropy in a wide temperature range.

\begin{figure}
\begin{center}
\resizebox{0.90\columnwidth}{!}{\includegraphics*{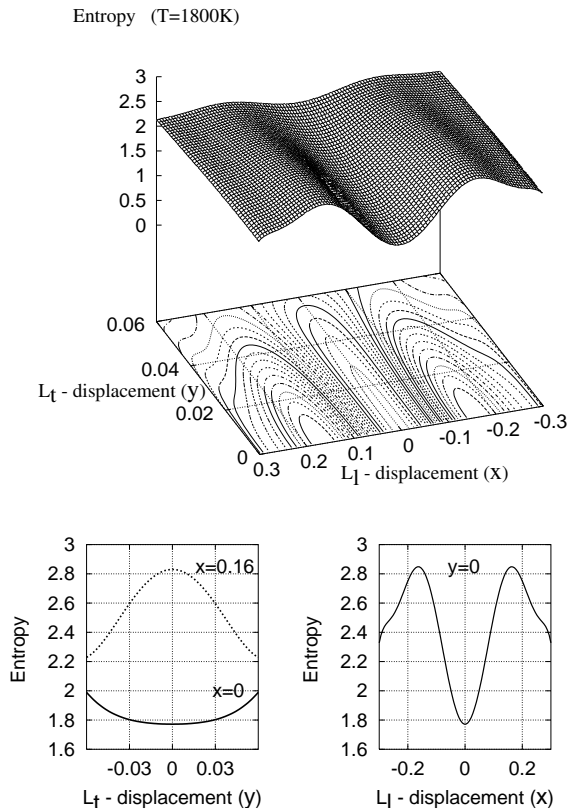}}
\caption{Electronic entropy at  $T=1800$K for the longitudinal and transverse atomic displacements corresponding to the vibrations with wave vector $k=\frac{2}{3}(1,1,1)$ in $\beta-$Zr. The displacements are given in units of the lattice parameter $a$.}
\label{fig:Pot+Entropy2}
\end{center}
\end{figure}

In Fig.~\ref{fig:Pot+Entropy2} the electronic entropy is calculated at $T=1800$K for the atomic displacements corresponding to both the longitudinal, $L_l$, and transverse, $L_t$, vibrations with wave  vector $k=\frac{2}{3}(1,1,1)$ in $\beta-$Zr. As seen from the figure, entropy is an intricate function of the displacements. The entropy minimum occurs at zero displacements corresponding to the atomic positions characteristic of the $\omega$  structure (see the right lower panel).

In the left lower panel of Fig.\ref{fig:Pot+Entropy2} the changes in entropy are shown at transverse displacements for the 
$\omega$ (solid line) and $bcc$ (dotted line) atomic configurations. The entropy is seen to increase for the atomic vibrations in the $\omega$ lattice and decrease in the $bcc$ one.

\section{Temperature dependence of entropy and free energy}   

\begin{figure}[!tbh]
\begin{center}
\resizebox{0.95\columnwidth}{!}{\includegraphics*{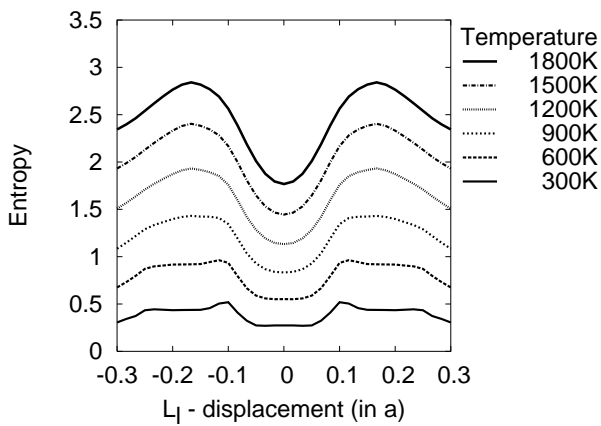}}
\caption{Entropy for the longitudinal displacements at different temperatures.}
\label{fig:T-free_entropy}
\end{center}
\end{figure}

As seen from Fig.~\ref{fig:Entrop_by_T}, the temperature dependence of the entropy differs for different atomic displacements $x,y$. 
In Fig. \ref{fig:T-free_entropy} the electronic entropy $S(x,0)$ is shown for various temperatures. The difference in entropy between the $\omega$ and $bcc$ lattices is seen to increase with temperature, the entropy maximum shifting from   $x=0.10$ to $x=0.2$.

\begin{figure}[!tbh]
\begin{center}
\resizebox{0.95\columnwidth}{!}{\includegraphics*{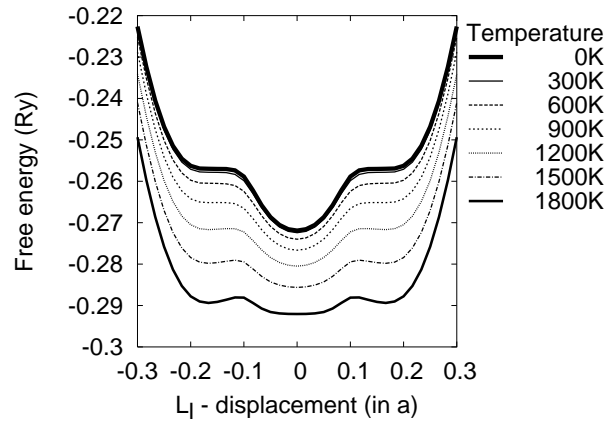}}
\caption{ Free energy for the longitudinal displacements at different temperatures.}
\label{fig:T-free_energy3}
\end{center}
\end{figure}

As the entropy enters in the free energy (\ref{Free_energy}) with negative sign, such a change in entropy must lead to a decrease of the difference in free energy between the  $\omega$ and $\beta$ phases with increasing temperature, and, as a consequence, to the variation with temperature of  the effective potential shape. 
In the 2D case the free energy as a function of the displacements $x,y$ has a shape similar to that in Fig.\ref{fig:Pot+Entropy}. Note that for each fixed temperature we obtain a particular effective potential  $W_T(x,y)$. 
As an illustrative example, we show in Fig.~\ref{fig:T-free_energy3} only the 1D potential $W_T(x,0)$ for the $L_l$ mode at different temperatures, as being the most important for further discussion.  
It can be seen that, at high temperatures, allowance for the electronic entropy results in a significant change of the effective potential. Although the potential remains strongly anharmonic at high temperatures, the energies of the $bcc$ and $\omega$ lattices become comparable.         

Thus, by calculating the free energy at different temperatures, we obtained the temperature-dependent effective potetials for the interacting longitudinal and transverse vibrations of $\beta$ zirconium. In general, the magnitude of the potential barrier between the $\omega$ and $\beta$ phases decreases with increasing temperature, which should affect both the vibration frequency and the interaction between the  $L_L$ and  $L_T$ modes.  

\section{Equations of motion for a particle in a potential field with random forces}
Let us consider the dynamics of motion of a pseudoparticle in the 2D effective potential $W(x,y)$ shown in Fig.\ref{fig:Pot+Entropy}. As follows from  the construction of $W(x,y)$, the variables $x,y$ are generalized collective variables corresponding to two waves of atomic displacements with the same wave vector $k=\frac{2}{3}(1,1,1)$ and different polarizations (longitudinal, $x$, and transverse, $y$). We assume that the effect of all the other phonons of the system may be represented as random forces acting on a system of two interacting oscillators. The dynamics of motion of two interacting vibrational modes embedded in a thermostat with random forces is defined by the stochastic differentional equations of the Langevin type

\begin{eqnarray}
\frac{d^2 x}{d\,t^2}+ \frac{\partial W(x,y)}{\partial x}+\gamma_x \frac{d\,x}
{d\,t}&=& F_x(t)\nonumber\\
\frac{d^2 y}{d\,t^2}+ \frac{\partial W(x,y)}{\partial y}+\gamma_y \frac{d\,y}
{d\,t}&=& F_y(t).
\label{Eq:Stoch_eq}
\end{eqnarray}
$ F_x(t),F_y(t)$ are random forces with correlators
\begin{eqnarray}
<F_i(t)>&=&0\label{Eq:White_noise1}\\
<F_i(t)F_j(t')>&=&2T \, \gamma\, \delta_{ij}\, \delta(t-t'),
\label{Eq:White_noise2}
\end{eqnarray}
$\gamma$ are coefficients of the vibration damping,
$T$ is the temperature of the thermostat.
Assuming all the higher momenta to be zero at $n \geqslant 3$
\begin{equation}
<F(t_1)F(t_2)...F(t_n)> =0,
\end{equation}
we obtain the standard definition of the random process $F(t)$ called {\it Gaussian white noise}.

The coefficient in (\ref{Eq:White_noise2}) was chosen so that Eqs.(\ref{Eq:Stoch_eq}) describe, in the limit of  large times, the relaxation of the distribution function  to the stationary Boltzmann distribution with temperature $T$  \begin{equation}
P({\bf X,\bar v},t\to\infty)=\exp\left[-\frac{m{\bar v}^2/2+W({\bf X})}{2k_bT}\right],
\label{Eq:bolzman}
\end{equation}
where $\bf X(t)$ is a  dynamical vector variable,
and the mean-square velocity  $\bar v$ in the limit $t\to\infty$ is equal to the thermodynamically equilibrium value at temperature $T$
\begin{equation}
\bar v^2=\lim_{t\to\infty} <v^2>=k_bT.
\end{equation}

The equilibrium distribution   (\ref{Eq:bolzman}) results from the action of two opposing tendencies. Owing to the presence of random forces ${\bf F}$ the velocity ${\bf V}$ 
(and, hence,  {\bf X}) becomes a random value, while the term describing the damping $\gamma {\bf V}$
suppresses ${\bf V}$, tending to reduce it to zero.
The set of equations (\ref{Eq:Stoch_eq}) was solved by numerical integration of stochastic differential equations, using a method \cite{Greenside}  being a generalization of the Runge-Kutta scheme to the stochastic differential equations. We used a 4-step method of third order with the parameters from Ref.~\onlinecite{Greenside}. 

\section{Spectral density}
\label{Sec:SP}

When solving the set of stochastic differential equations, a particular effective potential $W(T)$ was  constructed for each fixed temperature $T$ in  (\ref{Eq:White_noise2}). 
The calculation was performed with a time step $\Delta t=1.57 \cdot 10^{-16}sec.$, the number of steps in the realization being  $N_{sh}=3 \cdot10^8$. Thus, the total time of modelling was $t_{r}\approx 5 \cdot 10^{-8}sec.$, which is vastly greater than the period of vibrations of the chosen modes. The test calculations for  $N_{sh}=9 \cdot10^8$ have shown that the result remains uchanged with such an increase of the modelling time. 

The stochastic dynamical variables (coordinates $ X,Y$ and velocities $ V_X,V_Y$) found by solving Eqs.(\ref{Eq:Stoch_eq}) were used to calculate the autocorrelation velocity functions $\mathbb K_i(\tau)=<<V_i(0)V_i(\tau)>>$.  To this end the whole interval of modelling time was divided into realizations. The values of dynamical variables  $\bf  X(t_0),Y(t_0)$ obtained at the end of each realization were used as the starting values for the next one. In the average, the length of one realization was $N_r=100000$ steps ($1.57 \cdot 10^{-11}$sec). When calculating the correlators, the total number of realizations over which the averaging was performed  amounted to 3000.
  
\begin{figure}[!tbh]
\begin{center}
\resizebox{0.95\columnwidth}{!}{\includegraphics*{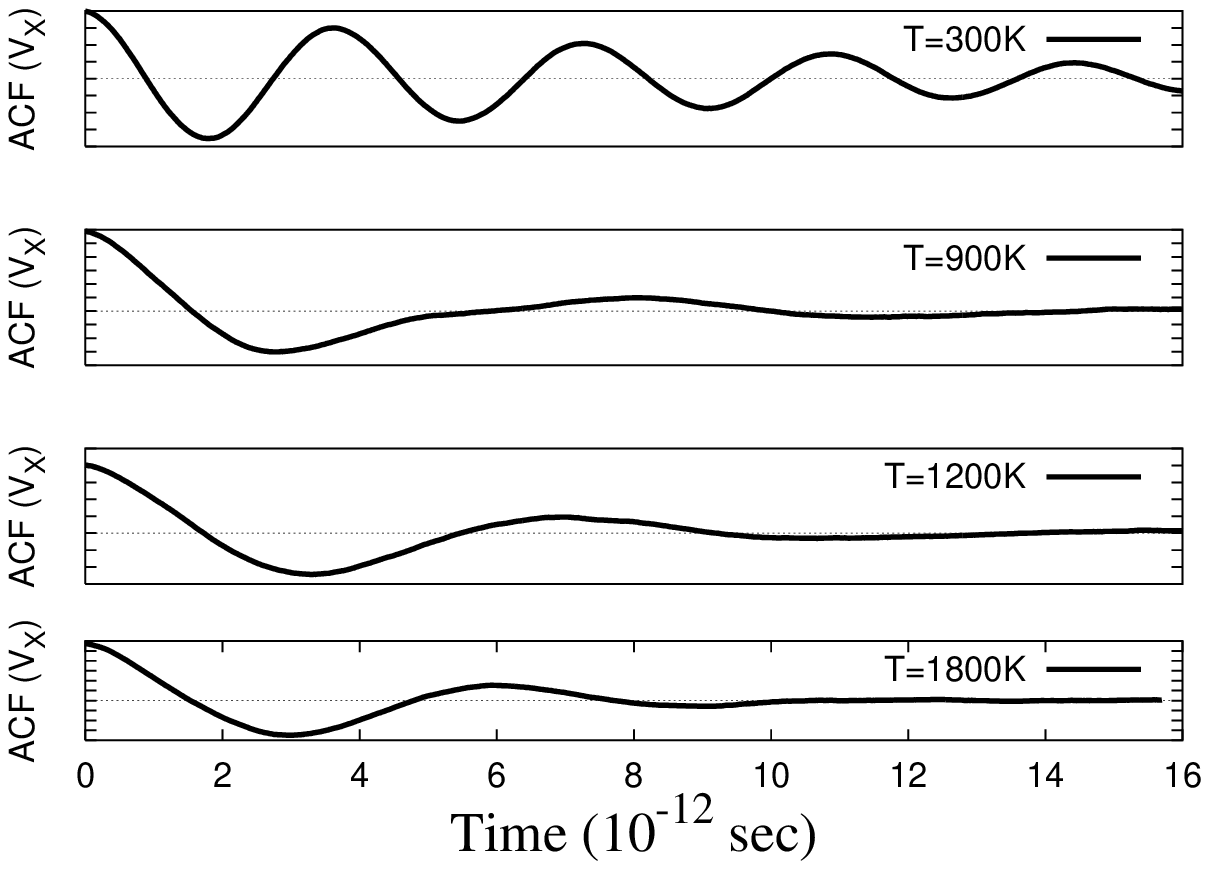}} \\
\resizebox{0.95\columnwidth}{!}{\includegraphics*{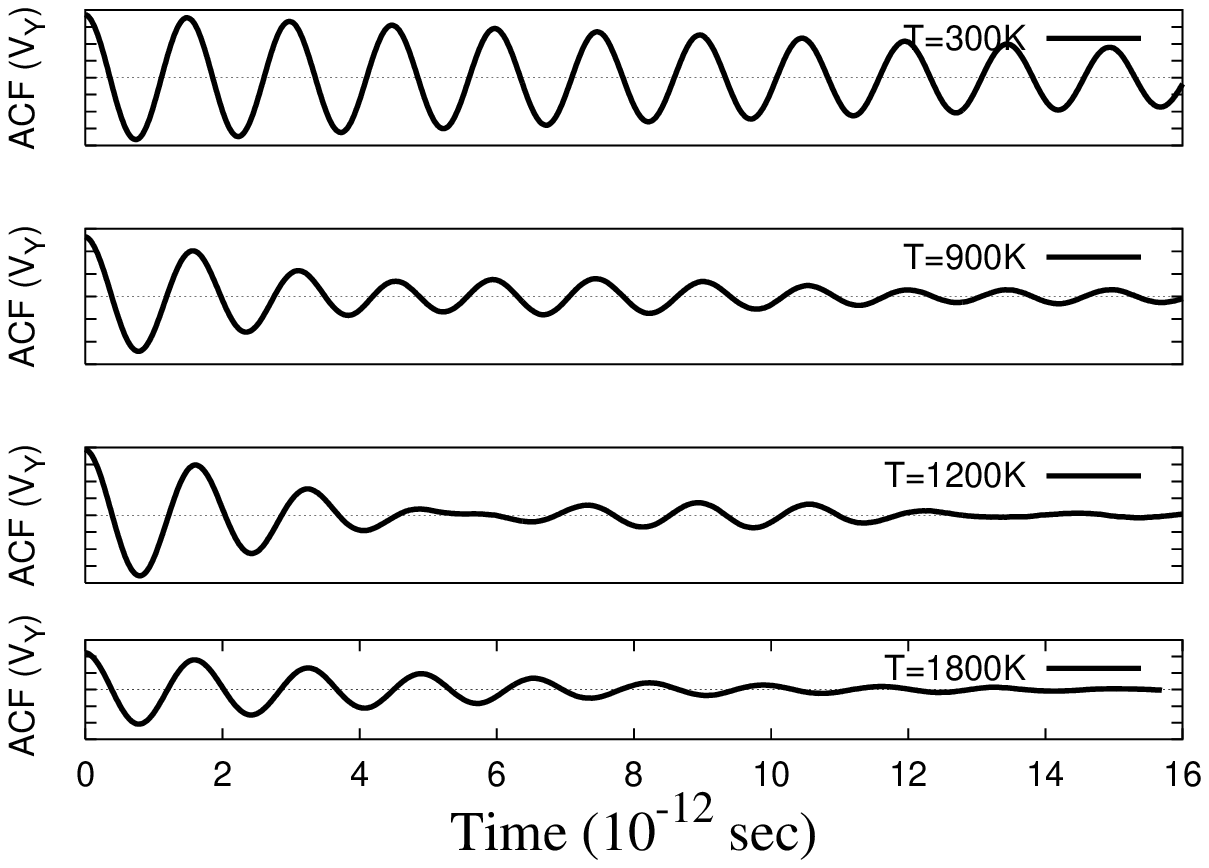}}
\caption{Autocorrelation velocity functions $<V_x(0)V_x(t))>$ (upper diagrams) and $<V_y(0)V_y(t))>$ (lower diagrams) at different temperatures T.} 
 \label{fig:AKF}
\end{center}
\end{figure}

  The autocorrelation velocity functions are shown in Fig.\ref{fig:AKF} for different temperatures.  
It can be seen that at low temperatures the autocorrelation functions oscillate for a time that exceeds the period of atomic vibrations, whereas at high temperatures the correlations decline rapidly. This means that at high temperatures the system quickly forgets about its initial state and comes into a  stationary one. The temperature dependence of the correlation time is determined by the conditions imposed on random external forces $F(t)$ acting on the system of two oscillators  (see relations (\ref{Eq:White_noise1},\ref{Eq:White_noise2})). 

The first relation reflects the fact that the mean force acting on the system of oscillators is equal to zero.
The second condition implies that the interaction with the thermostat is practically instantaneous, and the successive interactions do not correlate with one another. As seen from Eq.(\ref{Eq:White_noise2}), the magnitude of random external forces increases with temperature, which results in a  faster damping of the correlators.  It follows from Fig.\ref{fig:AKF} that at low ($T=300$K) and high($T=1800$K) temperatures the autocorrelation functions have practically the same oscillation frequency, while at intermediate temperatures there are additional oscillations. This is especially noticeable for the autocorrelation of the $y$ component of the velocity at $T=1200$K, and for $<V_x(t)V_x(t_0)>$ at $T=900$K.

\begin{figure}[!tbh]
\begin{center}
\resizebox{0.95\columnwidth}{!}{\includegraphics*{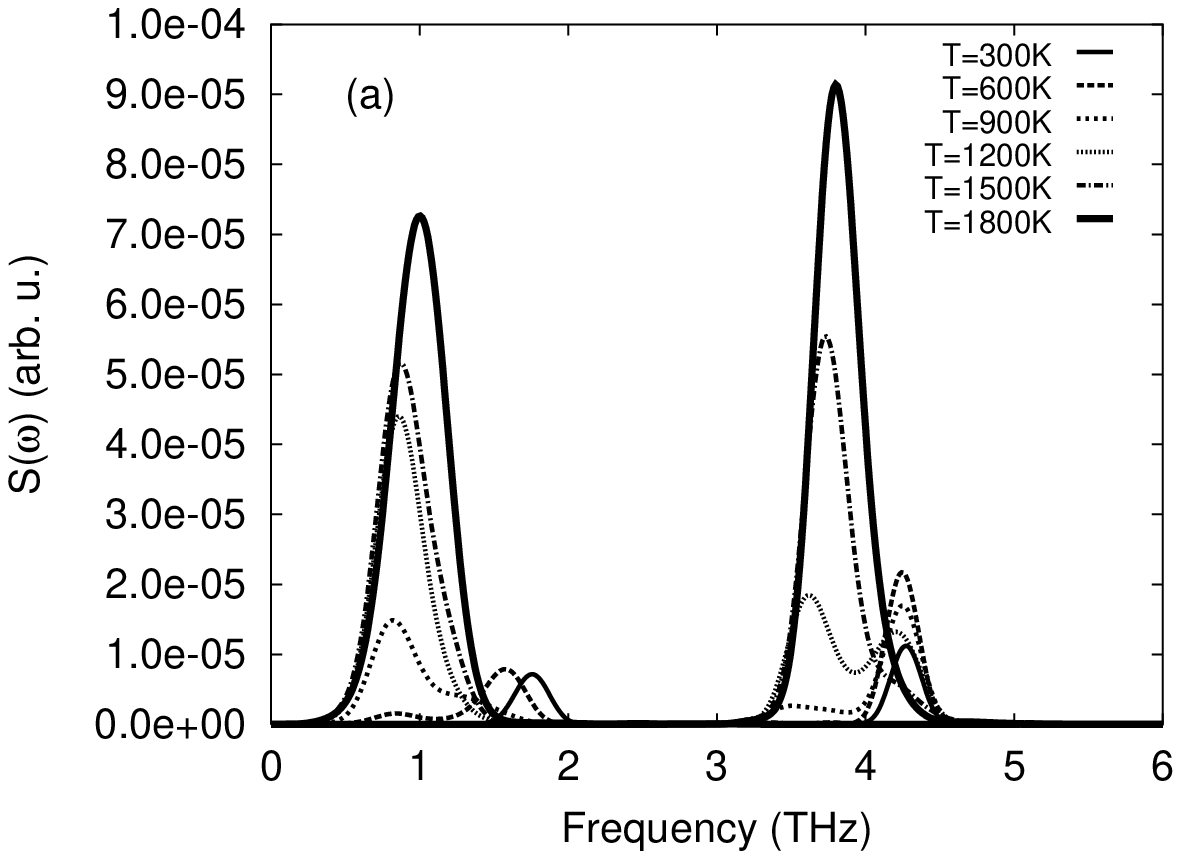}}
\resizebox{0.95\columnwidth}{!}{\includegraphics*{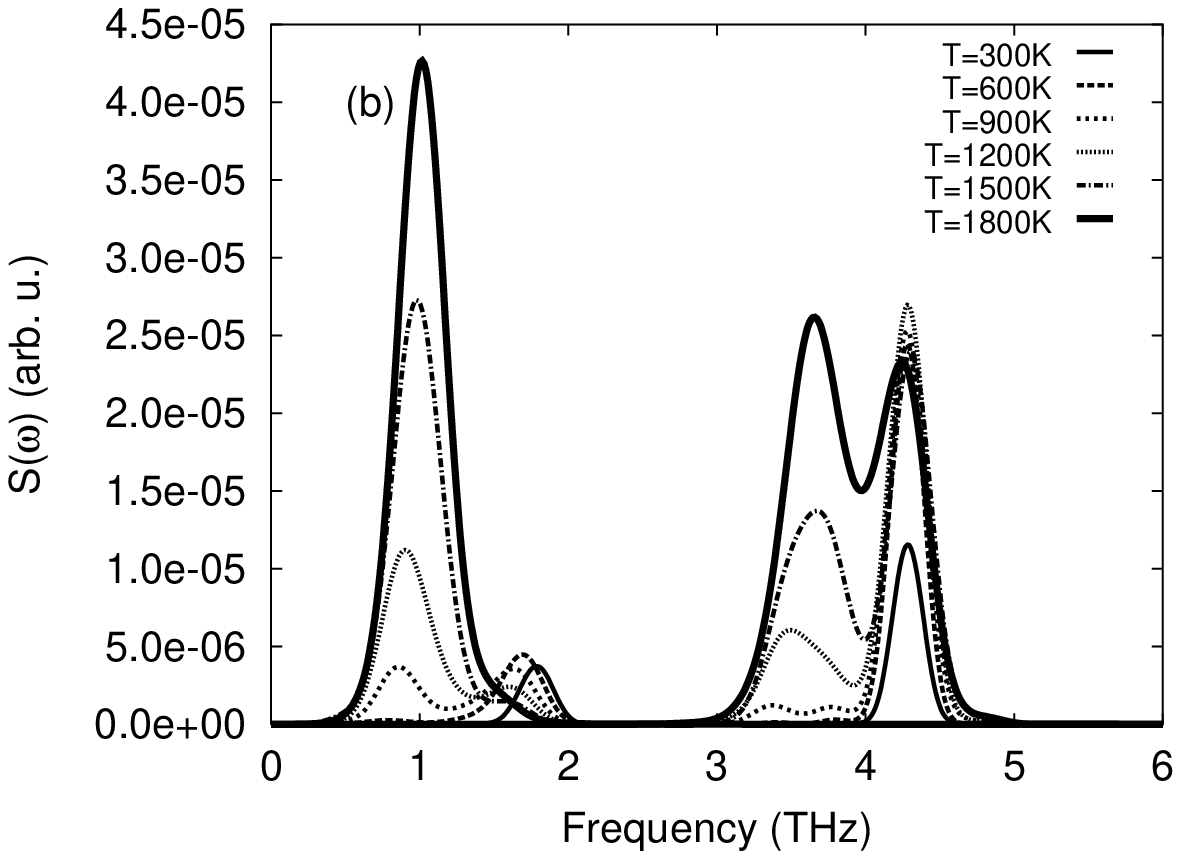}}
\caption{Spectral density of the $L-$phonon calculated for the temperature dependent (a) and temperature independent (b) effective potential.}
 \label{fig:Spec_density}
\end{center}
\end{figure}

The spectral vibration density $S_{i}(\omega)$ found from the autocorrelation velocity function $\mathbb K_{i}(\tau)$ \begin{equation}
S_{i}(\omega)=\frac{2}{\pi}\int\limits_0^{\infty}\cos(\omega \tau)\mathbb K_{i}(\tau)\,d\,\tau
\end{equation}
 is presented in Fig. \ref{fig:Spec_density}(a) for different temperatures. 
The first maximum of the spectral density at frequencies close to $\omega \approx 1 $THz is due to the longitudinal vibrations, and the second one, near $\omega \approx 4 $THz is connected to the transverse vibrations. As seen from the figure, the temperature variation leads to a significant change in the spectral density of  both longitudinal and transverse virations.

We first consider the fine structure of the spectral density of {\em longitudinal } vibrations, $S_x(\omega)$, and its evolution with temperature. At low temperatures there is only one peak of frequency $\omega_1^x \approx 1.9 $THz, which corresponds to the atomic vibrations near the sites appropriate to the $\omega$ lattice ($x=0$). With increasing temperature, the frequency of such vibrations diminishes, which is due to  a slight anharmonicity of the effective potential (Fig.\ref{fig:T-free_energy3}) near the origin. At temperature $T=600$K, overbarrier vibrations arise in the system,  leading to the appearance of an additional peak in the vicinity of $\omega_2^x \approx 0.9$THz in the $S_x(\omega)$ curve. With a rise in temperature, the probability of overbarrier vibrations increases, and at $T=900$K the share of such vibrations becomes dominant. This results in an increase of the spectral density intensity $S_x(\omega)$ at $\omega \approx 1.0$ THz and in its  decrease at $\omega > 1.5$ THz. This tendency persists with further increase in temperature, the maximum of the spectral density shifting to greater frequencies.  As seen in Fig.\ref{fig:Spec_density}(a), at $T=1800$K practically all $L_l$ vibrations are the overbarrier ones. 

An analogous change in the spectral density is also observed for the  {\em transverse} atomic displacements (the frequency range $ \omega \approx 3-4.5$ THz). 
Again, at room temperature we see one maximum of frequency $ \omega_1^y \approx 4.2$ THz, corresponding to the vibrations near  the sites of the $\omega$ lattice.
Then, with increasing temperature, an additional peak appears in the range  $\omega_2^y \approx 3.5$ THz. With further rise in temperature this peak shifts to higher frequencies and becomes the main one.

\begin{figure}[!tbh]
\begin{center}
\resizebox{0.95\columnwidth}{!}{\includegraphics*{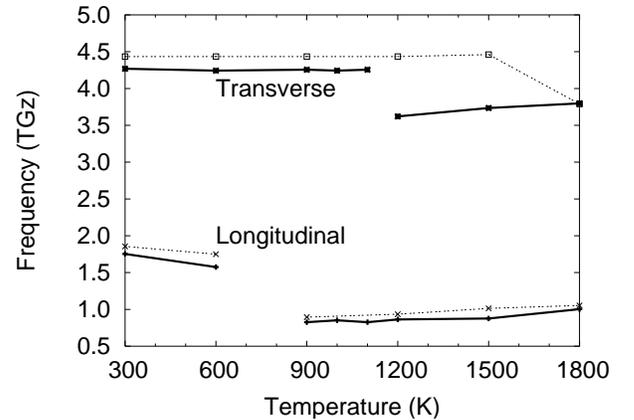}}
\caption{The temperature dependence of the frequency  $\omega_{max}$. The temperature-dependent effective potential is shown by the solid line, the temperature-independent one by the dashed line.}
 \label{fig:Omega_max}
\end{center}
\end{figure}      

In Fig.\ref{fig:Omega_max} the solid line shows the temperature dependence of the vibratonal frequency at which the spectral densities $S_x(\omega)$ and $S_y(\omega)$ have a maximum value (individual values of $\omega_{max}$ were estimated for the longitudinal and transverse components of $S(\omega))$. 
As seen, for both types of vibrations there is a discontinuity in $\omega_{max}$. From the above it follows that the discontinuity occurs at $T_{tr}$, at which the intensities are equal: $S_x(\omega_1^x)=S_x(\omega_2^x)$ or $S_y(\omega_1^y)=S_y(\omega_2^y)$. 

For the longitudinal vibrations the temperature $T_{tr}^x$ lies in the range from 600 to 900~Ê, and for the transverse ones $T_{tr}^y \approx 1200$~Ê. The fact that the temperatures $T_{tr}^i$ differ is due to different nature of the two-hump structure of $S_x(\omega)$ and $S_y(\omega)$. 
In the former case, the presence of a structure with similar intensities of the spectral density reflects the fact that the  overbarrier vibrations equal in number the vibrations near the $\omega$ centres, while in the latter case the similar intensities indicate that the shares of vibrations near the $bcc$ and $\omega$ sites are equal. Therefore, the spectral density of the longitudinal vibrations, $S_x(\omega)$, carries no immediate information about the proportion of time the system spends respectively in the $bcc$ and the $\omega$ lattice. Such an information is contained in the spectral density of the  vibrations, $S_y(\omega)$. It is seen from the above that at a temperature close to 1200Ê the probabilities of finding the system in the $\omega$ and the $bcc$ phase are equal. 
The temperature $T_{tr}^y$ can be considered  as the temperature of equilibrium between the two phases with respect to the atomic displacements characteristic of the longitudinal $L_l$ vibrations with wave vector $k=2/3(111)$. 

For reference, in Fig.\ref{fig:Spec_density}(b) is plotted the spectral density of vibrations calculated for a {\it temperature-independent} effective potential. In this case for all temperatures the effective potential was the same, equal to the potential found from the total energy without allowance for  entropy. A comparison between Fig.\ref{fig:Spec_density}(a) and Fig.\ref{fig:Spec_density}(b) shows that the greatest discrepancy is observed for the transverse vibrations. Namely, the two-hump structure of the spectral density of transverse vibrations persists up to $T=1800K$, which far exceeds the phase transition temperature in zirconium. And at $T=1200$K the main contribution to the spectral density comes from the vibrations with frequency $\omega \approx 4.2$THz, corresponding to the vibrations near the equilibrium position of the $\omega$ structure. The calculated values of $\omega_{max}$ for this case are shown in Fig.\ref{fig:Omega_max} by a dashed line. As seen from the figure, taking account of the temperature dependence of the effective potential leads to a decrease in the frequency of both longitudinal and transverse vibrations in the system and to a fall in temperature $T_{tr}^y$.

Thus, the calculation performed shows that the allowance made for the electronic entropy when modelling strongly anharmonic lattice vibrations results in a decrease of the calculated temperature value at which the $\beta$ phase of zirconium becomes stable to the atomic displacements corresponding to the $L$ phonon.  

\section{Conclusion} 
In this work we have used $\beta-Zr$ to discuss the  effect of temperature on the effective potential $W(x,y)$ and the spectral density of vibrations of the longitudinal, $L_l$, and transverse, $L_t$, interacting modes with wave vector ${\bf k}=2/3(111)$.
The effective potential acting on these modes at zero temperature was calculated in the frozen-phonon approximation within the electron density functional theory by the {\em{FP LMTO}} method \cite{SAVRAS,SAVR}.
For nonzero temperatures, the potential $W_T$ was defined as the difference of the free energies $F_T$ calculated for different atomic displacements, $W_T(x,y)=F_T(x,y)-F_T(0,0)$. The required electronic entropy was defined through the density of electron states and the Fermi-Dirac distribution function. 

The results obtained show that in zirconium in the temperature range up to 2000Ê the entropy may be well approximated by a linear function of the density of electronic states. The shape of the effective potential $W(x,y)$ for the $L_l$ and $L_t$ vibrations strongly depends on the temperature. With increasing temperature the height of the energy barrier between the $bcc$ and $\omega$ structures decreases. As a result, the $bcc$ lattice of Zr becomes stable with respect to these modes at lower temperatures. An analysis of the vibrational frequency of the {\em transverse} $L_t$ mode  showed the temperature, at which the $\beta$ phase of Zr becomes unstable with respect to the longitudinal $L_l$ displacements, to be $T_L = 1150 \pm 25$K. This value practically coincides with the experimental temperature of the $\beta \to \alpha$ transition: $T_{\beta \to \alpha} = 1136$K \cite{Tonkov}. Thus, the $bcc$ lattice of Zr is dynamically stable with respect to the longitudinal $L_l$ vibrations owing to the intrinsic anharmonicity of this mode.  Of considerable importance is the electronic entropy which substantially enlarges the stability region of $\beta-Zr$ with a decrease in temperature.
It is commonly believed today that under atmospheric pressure and low temperatures the structural instability of $\beta$-Zr is due to the softening of the transverse phonon with wave vector ${\bf k}=\frac{1}{2}(110)$ ($N_t$-phonon) \cite{Heiming,Ostanin-2000}. The result obtained in this paper points to the fact that at zero pressure there is at least one more process responsible for  the structural instability of $bcc$ zirconium: the atomic displacements corresponding to the longitudinal $L_l$ phonon. This brings up the question as to whether the coincidence of the temperatures $T_{tr}^y$ and $T_{\beta \to \alpha}$ is accidental, or  the $L_l$ mode is really of considerable, if not decisive, importance in the  $\beta \to \alpha$ structural transformation in zirconium.

 \begin{acknowledgments}
 The author acknowledges the support from the  RFBR Grant No. 04-02-16680
\end{acknowledgments}


\begin{thebibliography} {99}
\bibitem {Tonkov} E.Yu.~Tonkov, {\it High Pressure Phase Transformations} 
(Gordon and Breach Science Publishers, Philadelphia, 1992)   Vol. 2. 
\bibitem {Heiming1} A.~Heiming, W.~Petry, J.~Trampenau, M.~Alba, C.~Herzig, G.~Vogl,
Phys. Rev. B {\bf 40}, 11425 (1989).
\bibitem {Heiming}A.~Heiming, W.~Petry, J.~Trampenau, M.~Alba, C.~Herzig, H.R.~Schober, G.~Vogl, Phys. Rev. B {\bf 43}, 10948 (1991).   
\bibitem{Chen-85} Y.~Chen, C.-L.~Fu, K.-M.~Ho, and B.N.~Harmond, Phys. Rev. B {\bf 31},  6775 (1985).
\bibitem {Salamatov-96}E.I.~Salamatov, Phys. Stat. Sol. {\bf 197}, 323 (1996).   
\bibitem {Ostanin-98}S.A.~Ostanin, E.I.~Salamatov, and V.Yu.~Trubitsin, Phys. Rev. B {\bf 57}, 5002 (1998).
\bibitem {Ho-84}K.-M.~Ho, C.L.~Fu, and B.N.~Harmon, Phys. Rev. B {\bf 29}, 1575 (1984).
\bibitem {Akahama1} Y.~Akahama,  M.~Kobayashi, and H.~Kawamura,
J. Phys. Soc. Japan, {\bf 59}, {\it 11}, 3843  (1990).
\bibitem {Akahama2} Y.~Akahama,  M.~Kobayashi, and H.~Kawamura,
J. Phys. Soc. Japan, {\bf 60}, {\it 10}, 3211  (1991).     
\bibitem {Trubitsin-2004}V.~Trubitsin, S.~Ostanin, Phys. Rev. Lett. {\bf 98}, 385503-1 (2004).
\bibitem {EWW}  O.~Eriksson, J.M.~Wills, D.~Wallace,  Phys. Rev. B $\bf{46}$, 5220 (1992).  
\bibitem {Gornostyr-96}Yu.N.~Gornostyrev and M.I.~Katsnelson, A.V.~Trefilov, S.V.~Tret'jakov, Phys. Rev. B {\bf 56}, 3286 (1895).    
\bibitem {SAVRAS}    S.Yu.~Savrasov and D.Yu.~Savsasov,   Phys. Rev. B  $\bf 46$, 12181 (1992).
\bibitem {SAVR}  S.Y.~Savrasov, Phys.Rev.B, {\bf 54}, 1640 (1996).
\bibitem {GGA96}  J.P.~Perdew, K~Burke, M~Emzerhof, Phys. Rev. Lett.B $\bf 77$, 3865 (1996).    
\bibitem {Pinsook-98}U.~Pinsook and G.J.~Ackland, Phys. Rev. B {\bf 58}, 11252 (1998).
\bibitem {Pinsook-99}U.~Pinsook and G.J.~Ackland, Phys. Rev. B {\bf 59}, 13642 (1999).
\bibitem {Gornostyr-99}Yu.N.~Gornostyrev, M.I.~Katsnelson, A.R.~Kuznetsov, A.V.~Trefilov, Pis'ma v JETF {\bf 70}, {\it  6}, 376 (1999).   
\bibitem {Ostanin-2000}S.A.~Ostanin, E.I.~Salamatov, and V.Yu.~Trubitsin, High Pressure Research {\bf 17}, 385 (2000).
\bibitem {Olijnyk}H.~Olijnyk and A.P.~Jephcoat, Phys. Rev. B {\bf 56}, 10751 (1997).
\bibitem {Greenside}H.S.~Greenside and E.~Helfand, Bell Syst.Tech.J. {\bf 60}, 1927 (1981).
\bibitem {Lifshic} I.M.~Lifshits, M.Ya.~Azbel', M.I.~Kaganov,  {\it Electron theory of metals} (Nauka, Moscow 1971).  
\bibitem {Dubos}  O.~Dubos, W.~Petry, J.~Neuhaus, and B.~Hennion, Eur.Phys.J. B $\bf 3$, 447 (1998).
\bibitem {Olijnyk2}  H.~Olijnyk, J.Phys.:Condens Matter $\bf 11$, 6589 (1999).
\bibitem {Althoff}  J.D.~Althoff, P.B.~Allen, R.M.~Wentzcovitch, and J.A.~Moriarty,   Phys.Rev.B., $\bf 48$, 13253 (1993).
\bibitem {Kampen} N.G.~Van Kampen, Stochastic processes in physics and chemistry, North-Holland Physics Publishing (1984).

\end{thebibliography}
\end{document}